\documentclass[onecolumn,preprintnumbers,amsmath,amssymb,nofootinbib]{revtex4}

\usepackage{graphicx}
\usepackage{bm}
\usepackage{amsmath}

\newcommand{\be}{\begin{equation}}
\newcommand{\ee}{\end{equation}}
\newcommand{\bea}{\begin{eqnarray}}
\newcommand{\eea}{\end{eqnarray}}
\newcommand{\bdm}{\begin{displaymath}}
\newcommand{\edm}{\end{displaymath}}
\newcommand{\beas}{\begin{eqnarray*}}
\newcommand{\eeas}{\end{eqnarray*}}

\newcommand{\mD}{\mathcal{D}}

\newcommand{\av}[1]{\left< #1\right>}
\newcommand{\dom}[1]{#1_{\mathcal{D}}}
\newcommand{\intdom}[1]{\frac{1}{\dom{V}}\int_{\mathcal{D}} #1\sqrt{h}d^3\mathbf{x}}
\newcommand{\Hd}{\frac{\dom{\dot{a}}}{\dom{a}}}
\newcommand{\Rd}{\mathcal{R}_{\mathcal{D}}}
\newcommand{\Pd}{\mathcal{P}_{\mathcal{D}}}
\newcommand{\BPd}{\mathcal{\beta}^P_{\mathcal{D}}}
\newcommand{\Qd}{\mathcal{Q}_{\mathcal{D}}}
\newcommand{\BQd}{\mathcal{\beta}^Q_{\mathcal{D}}}
\newcommand{\Fd}{\mathcal{F}_{\mathcal{D}}}
\newcommand{\Td}{\mathcal{T}_{\mathcal{D}}}
\newcommand{\Sd}{\mathcal{S}_{\mathcal{D}}}
\newcommand{\Ld}{\mathcal{L}_{\mathcal{D}}}
\newcommand{\Vd}{\dom{V}}

\newcommand{\bkr}{\overline{\rho}}
\newcommand{\bkp}{\overline{p}}

\newcommand{\weff}{w_\mathrm{eff}}
\newcommand{\pd}{\partial}
\newcommand{\hub}{\frac{\dot{a}}{a}}

\begin{document}

\preprint{HD-THEP-09-03}

\title{Gauges and Cosmological Backreaction}

\author{Iain A. Brown$^1$}
\email{I.Brown@thphys.uni-heidelberg.de}
\author{Juliane Behrend$^2$}
\email{Juliane.Behrend@uni-ulm.de}
\author{Karim A. Malik$^3$}
\email{K.Malik@qmul.ac.uk}
\affiliation{1: Institut f\"ur Theoretische Physik, Philosophenweg 16, 69120 Heidelberg, Germany}
\affiliation{2: Institut f\"ur Theoretische Physik, Albert-Einstein-Allee 11, 89069 Ulm, Germany}
\affiliation{3: Astronomy Unit, School of Mathematical Sciences, Queen Mary University of London, Mile End Road, London, E1 4NS, United Kingdom}

\date{\today}

\begin{abstract}
We present a formalism for spatial averaging in cosmology applicable to general spacetimes and coordinates, and allowing the easy incorporation of a wide variety of matter sources. We apply this formalism to a Friedmann-Lema\^itre-Robertson-Walker universe perturbed to second-order and present the corrections to the background in an unfixed gauge. We then present the corrections that arise in uniform curvature and conformal Newtonian gauges.
\end{abstract}

%\pacs{04.25.Nx, 95.36.+x, 98.80.-k, 98.80.Jk}

\maketitle

\section{Introduction}
The issue of the averaging, or fitting, problem in cosmology is an old one and addresses the question of whether a Friedmann-Lema\^itre-Robertson-Walker (FLRW) metric is a reasonable approximation to the metric of the universe on large scales, and whether the evolution of the universe follows the usual FLRW dynamics. As the Einstein tensor is nonlinear, even should the metric of the universe be of FLRW form there is no guarantee that the same is true for the dynamics, and corrections therefore arise that act as an effective fluid. This effect is typically known as cosmological backreaction. The study of averaging in cosmology dates back at least to the 1960s \cite{Shirokov63} and was revisited in subsequent decades (see for example \cite{Ellis84,Futamase89,Kasai92,BuchertEhlers95,Futamase96,Zalaletdinov96} for an inexhaustive selection of studies). Following the discovery of the ``dark energy'' in the late 1990s there was a great increase of study in this area (e.g. \cite{Russ97,Buchert99,Buchert01,Wetterich03,Rasanen03}). Recent studies have begun to quantify the effect in perturbed FLRW systems with varying results; see for example \cite{Li07,Vanderveld07,Rasanen08,Li08,Paranjape08} and for further references see \cite{Brown09}. However, many of these studies were limited to pure dust (``Einstein-de Sitter'') universes, and frequently to spatial slicings in synchronous coordinates. In previous work \cite{Behrend08} (hereafter Paper I) and \cite{Brown09} (hereafter Paper II), as in for example \cite{Kasai06,Tanaka06}, a Newtonian slicing was employed instead. Paper II additionally introduced a general formalism applicable to systems with a variety of fluids for a wide range of redshifts.

When considering the issue of backreaction in perturbation theory, the question of gauges has to be carefully considered. %The results of, for example, \cite{Li07} and Paper II cannot necessarily be directly compared, as the former were calculated in conformal synchronous gauge and the latter in Newtonian gauge, which introduces complications in the definition of the spatial volume. 
The use of different gauges in backreaction introduces not least ambiguities in the definition of the spatial volume, which we consider particularly in flat gauge. Furthermore, it is not immediately obvious that there does not exist a gauge that is particularly suited to studies of backreaction -- in which the correction terms take their simplest forms or, indeed, could be shown to vanish identically. (As an example, a flat gauge would remove the curvature correction, while a Newtonian gauge would simplify the system by removing anisotropic terms.)

As yet it has been impossible to study this issue in its full generality, as to consider general gauges it is necessary to possess a formalism that incorporates a shift vector. More generally, a formalism that includes a shift vector can consider spacetimes in general coordinates, rather than those considering a particular symmetry. The central aim of this paper is to develop this formalism, and as a first example, apply it to perturbed Robertson-Walker universes in an unfixed gauge.\footnote{The issue of gauges in backreaction was considered in a study with somewhat different aims in \cite{Gasperini09}, while a formalism dealing with general coordinates was presented in \cite{Larena09} while the first version of this work was in final preparation.} It is then a simple matter to choose two promising gauges -- the uniform curvature, or flat, and the conformal Newtonian gauges -- and reduce the general expressions to particular cases. We immediately find that, unsurprisingly, the backreaction cannot be removed through a judicious choice of gauge. Moreover, these two obvious choices remain relatively complicated, although uniform curvature gauge is arguably physically clearer when considering spatial volumes and Fourier space, while Newtonian gauge yields simpler forms. In \S\ref{Average} we present a brief overview of the 3+1 separation before generalising the familiar spatial averaging procedure to the most general case. We apply this general formalism to a linearly-perturbed FLRW universe in an unfixed gauge in \S\ref{PertTheory} and present the backreaction quantities in full detail before considering the specialisation to flat and conformal Newtonian gauges in \S\ref{Gauges}. We end with a brief discussion.

\section{Generalised Averaging in Cosmology}
\label{Average}
The standard spatial averaging procedure in cosmology is based on a 3+1 split, where spacetime is foliated with purely spatial hypersurfaces upon which an averaging domain is defined. Assuming global hyperbolicity and a metric signature $(-+++)$, a spacetime can be foliated with a family of 3-surfaces described by coordinates $x^\mu=(t,x^a)$ and with a normal 4-vector
\be
  n^\mu=\frac{1}{\alpha}(1,-\beta^i), \quad n^\mu n_\mu=-1 .
\ee 
Here the threading between corresponding events on neighbouring surfaces is defined by the lapse function $\alpha$ and the shift vector $\beta^i$. The projection tensor onto the 3-surfaces is
\be
  h_{\mu\nu}=g_{\mu\nu}+n_\mu n_\nu 
\ee
and the induced 3-metric is
\be
  h_{ij}=g_{\mu\nu}h^\mu_ih^\nu_j .
\ee
The line element is then
\be
\label{3+1LineElement}
  ds^2=(-\alpha^2+\beta_i\beta^i)dt^2+2\beta_idtdx^i+h_{ij}dx^idx^j ,
\ee
where $\beta_i=h_{ij}\beta^j$.

The embedding of the hypersurfaces is described by the extrinsic curvature, which is given by the Lie derivative of the metric along the normal,
\be
\label{excurv}
  K_{ij}=-\frac{1}{2}\mathcal{L}_nh_{ij}.
\ee
This can be rewritten as
\be
  \alpha\mathcal{L}_nh_{ij}=\frac{\partial}{\partial t}h_{ij}-2\mD_{(i}\beta_{j)}=-2\alpha K_{ij}
\ee
where $\mD_i$ is the covariant derivative on the 3-surface and brackets denote symmetrisation on the enclosed indices.

Using $\varrho_{(a)}$, $J^{(a)}_i$ and $S^{(a)}_{\mu\nu}$ respectively to denote a fluid's energy density, current density and stress tensor on the 3-surface of a fluid species $a$, the 3+1 decomposition of the stress-energy tensor is
\be
\label{FluidsSurface}
  T^{(a)}_{\mu\nu}=\varrho_{(a)}n_\mu n_\nu+2n_{(\mu}J^{(a)}_{\nu)}+S^{(a)}_{\mu\nu}.
\ee
Here
\be
 \varrho_{(a)}=n^\mu n^\nu T^{(a)}_{\mu\nu}, \quad J^{(a)}_i=-n^\mu h^\nu_iT^{(a)}_{\mu\nu}=h^\nu_iJ^{(a)}_\nu,
  \quad S^{(a)}_{ij}=h_i^\mu h_j^\nu T^{(a)}_{\mu\nu}
\ee
and by construction $n^\mu J^{(a)}_{\mu}=n^\mu S^{(a)}_{\mu\nu}=0$. We retain $\rho_{(a)}$ and $p_{(a)}$ to denote the fluid rest-frame energy density and isotropic pressure respectively. Projecting the Einstein equations onto the hypersurface and along its normal results in the Hamilton constraint equation
\be
  \mathcal{R}+K^2-K^i_jK^j_i=16\pi G\sum_{a}\varrho_{(a)}+2\Lambda,
\ee
where $\mathcal{R}$ is the Ricci scalar on the surface and $K=K^i_i$, and the evolution equation
\bea
  \alpha\frac{\partial}{\partial t}K_{ij}&=&\alpha^2\left(\mathcal{R}_{ij}-2K^a_iK_{aj}+KK_{ij}-8\pi G\sum_{a}S^{(a)}_{ij}+4\pi Gh_{ij}\sum_{a}\left(S_{(a)}-\varrho_{(a)}\right)-\Lambda h_{ij}\right) \nonumber \\
    && \quad -\alpha\left(\mD_i\mD_j\alpha-\beta^a\mD_aK_{ij}-2K_{a(i}\mD_{j)}\beta^a\right) .
\eea
It also results in a momentum constraint equation with which we are not here concerned. For more details on the 3+1 split, we refer the reader to \cite{Gourgoulhon07,York79,Wald}.

The coordinate freedoms now allow us to eliminate four degrees of freedom. In previous studies of cosmological backreaction it has been standard to choose a synchronous slicing by setting $\alpha=1$, $\beta^i=0$ (e.g. \cite{Russ97,Buchert01,Li07}). An alternative has been to employ a ``Newtonian'' slicing by choosing $\beta^i=0$ and diagonalising the scalar sector of $h_{ij}$ (e.g. \cite{Kasai06,Tanaka06,Behrend08,Brown09}). In the following we do not fix our slicing and the coordinate freedoms are left implicit, allowing us to study spacetimes in arbitrary coordinate systems.

Now let
\be
  \dom{V}=\int_{\mD}\sqrt{h}d^3\mathbf{x}
\ee
be the definition of the volume of a finite domain $\mD$ lying on the inhomogeneous 3-surface where $h$ is the determinant of the 3-metric. Then the average of a scalar quantity $A$ in this domain can be defined as
\be
\label{AverageDef}
  \av{A}=\intdom{A} .
\ee
This averaging procedure cannot be sensibly applied to tensorial quantities since it is not invariant under a change of coordinates but retains its meaning for a scalar quantity. It has been used by most of the studies into cosmological averaging (e.g. \cite{Buchert99,Wetterich03,Rasanen03}); for a brief overview of alternatives see \cite{Stoeger99,Behrend08-2}. The averages are taken across spacelike hypersurfaces and therefore do not immediately relate to observation; for an approach at connecting a 3+1 averaging formalism with cosmology, see for example \cite{Larena08}. We can define an average ``Hubble rate'' from the expansion of the volume by
\be
  3\Hd=\frac{\dom{\dot{V}}}{\dom{V}}.
\ee
The derivative of the volume with respect to time is
\be
  \frac{\dom{\dot{V}}}{\dom{V}}=\intdom{\left(\frac{1}{2}h^{ij}\dot{h}_{ij}\right)}=\intdom{\left(-\alpha K+\mD^i\beta_i\right)}
\ee
where we have used that $\dot{h}=h^{ij}\dot{h}_{ij}$. The Hubble rate defined by the volume expansion is then
\be
  3\Hd=-\av{\alpha K+\mD^i\beta_i}
\ee
with $\mD^i=h^{ij}\mD_j$.\footnote{The definition of the Hubble rate is not unique and it is common to instead define it from the expansion scalar of a fluid. The relation between the approaches was recently calculated in \cite{Larena09}. We retain our present definition to avoid connecting our slicing directly with fluid quantities while leaving free the possibility of doing so later. This simplifies the presentation at the cost of physical clarity.} Integrating this yields the average scale factor $\dom{a}$. Combining the average Hubble rate with the averaged Hamiltonian constraint yields
\be
\label{AvHub1}
  \left(\Hd\right)^2=\frac{8\pi G}{3}\sum_a\av{\alpha^2\varrho_{(a)}}+\frac{1}{3}\av{\alpha^2}\Lambda-\frac{1}{6}\left(\Rd+\Qd^T\right)
\ee
which can be interpreted as a Friedmann equation for $\dom{a}$ in the domain, where
\be
  \Rd=\av{\alpha^2\mathcal{R}}, \quad
  \Qd^T=\av{\alpha^2\left(K^2-K^i_jK^j_i\right)}-\frac{2}{3}\av{\alpha K-\mD^i\beta_i}^2
\ee
are corrections from the averaged spatial curvature and the kinematical backreaction. The latter can be separated into the usual $\Qd$ (e.g. \cite{Russ97,Buchert99,Buchert01} and Paper II) which is related to the variance of the extrinsic curvature, and $\BQd$ which is dependent on the divergence of the shift,
\be
  \Qd=\av{\alpha^2\left(K^2-K^i_jK^j_i\right)}-\frac{2}{3}\av{\alpha K}^2, \quad
  \BQd=\frac{2}{3}\av{\mD^i\beta_i}\left(2\av{\alpha K}-\av{\mD^i\beta_i}\right), \quad
  \Qd^T=\Qd+\BQd.
\ee

Spatial averaging does not commute with differentiation with respect to time. The commutator between these operations applied to a scalar quantity $A$ is given by
\be
  \av{\dot{A}}=\av{A}^\cdot+\frac{\dom{\dot{V}}}{\dom{V}}\av{A}-\intdom{\left(\frac{1}{2}Ah^{ij}\dot{h}_{ij}\right)}
\ee
which can be written
\be
  \av{\dot{A}}=\av{A}^\cdot+\av{A\left(\alpha K-\mD^i\beta_i\right)}-\av{A}\av{\alpha K-\mD^i\beta_i} .
\ee

The trace of the evolution equation for the extrinsic curvature is
\be
\label{EvTrace}
  \alpha\dot{K}=\alpha^2\left(\mathcal{R}+K^2-12\pi G\sum_a\varrho_{(a)}+4\pi G\sum_aS_{(a)}-3\Lambda\right)-\alpha\mD^i\mD_i\alpha+\alpha\beta^a\mD_aK.
\ee
Since
\be
  \av{\alpha\dot{K}}=\av{\alpha K}^\cdot+\av{\alpha^2K^2}-\av{\alpha K}^2+\av{\alpha K}\av{\mD^i\beta_i}-\av{\alpha K\mD^i\beta_i}-\av{\dot{\alpha}K}
\ee
and
\be
  \av{\alpha K}^\cdot=-3\frac{\dom{\ddot{a}}}{\dom{a}}+3\left(\Hd\right)^2+\av{\mD^i\beta_i}^\cdot,
\ee
we have after some manipulation and the use of (\ref{AvHub1}) that the average of (\ref{EvTrace}) can be written
\be
\label{AvRay1}
  \frac{\dom{\ddot{a}}}{\dom{a}}=-\frac{4\pi G}{3}\sum_a\av{\alpha^2\left(\varrho_{(a)}+S_{(a)}\right)}+\frac{1}{3}\av{\alpha^2}\Lambda+\frac{1}{3}\left(\Qd^T+\Pd+\BPd\right) .
\ee
The dynamical backreaction \cite{Buchert02,Behrend08,Brown09} is
\be
  \Pd=\av{\alpha\mD^i\mD_i\alpha}-\av{\dot{\alpha}K},
\ee
and the component arising from the variation of the shift is
\be
  \BPd=\av{\left(\mD^i\beta_i\right)^\cdot+\left(\mD^i\beta_i\right)^2-2\alpha K\mD^i\beta_i-\alpha\beta^i\mD_iK} .
\ee
Equation (\ref{AvRay1}) can be interpreted as a ``Raychaudhuri equation'' for the scale factor $\dom{a}$.

The energy density and isotropic pressure are defined on the 3-surface in (\ref{FluidsSurface}). However, it is usual in cosmology to separate the stress-energy tensor with respect to the fluid's 4-velocity $u_{(a)}^\mu$,
\be
  T^{(a)}_{\mu\nu}=\rho_{(a)}u^{(a)}_\mu u^{(a)}_\nu+p_{(a)}\tilde{P}^{(a)}_{\mu\nu}+2q^{(a)}_{(\mu}u^{(a)}_{\nu)}+\pi^{(a)}_{\mu\nu} \iff \rho_{(a)}=u_{(a)}^\mu u_{(a)}^\nu T^{(a)}_{\mu\nu}, \quad 3p_{(a)}=\tilde{P}_{(a)}^{\mu\nu}T^{(a)}_{\mu\nu}
\ee
where
\be
  \tilde{P}^{\mu\nu}_{(a)}=h^{\mu\nu}+u^\mu_{(a)}u^\nu_{(a)}
\ee
projects tensors onto the 3-surface orthogonal to the 4-velocity, the 4-velocity is normalised to $u_{(a)}^\mu u^{(a)}_\mu=-1$, and $\rho_{(a)}$ and $p_{(a)}$ are the fluid's rest-frame energy density and isotropic pressure respectively. Let
\be
  \Fd^{(a)}=\frac{8\pi G}{3}\av{\left(h^{\mu\nu}-\tilde{P}_{(a)}^{\mu\nu}\right)T^{(a)}_{\mu\nu}}=\frac{8\pi G}{3}\av{\left(n^\mu n^\nu-u_{(a)}^\mu u_{(a)}^\nu\right)T^{(a)}_{\mu\nu}}
\ee
correct between the 3-surface and the fluid rest-frames. The effective, volume-averaged Friedmann and Raychaudhuri equations can therefore be written
\bea
\label{BuchertHub}
  \left(\Hd\right)^2&=&\frac{8\pi G}{3}\sum_a\av{\alpha^2\rho_{(a)}}+\frac{1}{3}\av{\alpha^2}\Lambda-\frac{1}{6}\left(\Rd+\Qd^T-6\sum_a\Fd^{(a)}\right), \\
\label{BuchertRay}
  \frac{\dom{\ddot{a}}}{\dom{a}}&=&-\frac{4\pi G}{3}\sum_a\av{\alpha^2\left(\rho_{(a)}+3p_{(a)}\right)}
   +\frac{1}{3}\av{\alpha^2}\Lambda+\frac{1}{3}\left(\Pd+\BPd+\Qd^T-3\sum_a\Fd^{(a)}\right) .
\eea
We therefore now have definitions of effective, averaged Friedmann and Raychaudhuri equations valid for any spacetime that allows a 3+1 split, with arbitrary lapse and shift, and an arbitrary number of fluids. Equations (\ref{BuchertHub}, \ref{BuchertRay}) are the main results of this work.\footnote{These equations can be compared with equations (36) and (37) of \cite{Larena09}.}

It is tempting to immediately define an effective energy density, pressure and equation of state of the ``backreaction fluid'' and such can certainly be done. Simply interpreting $\sum_a\av{\alpha^2\rho_{(a)}}$ and $\sum_a\av{\alpha^2p_{(a)}}$ as the average density and pressure in the domain, the effective energy density and pressure of the backreaction can be defined as
\be
\label{RawEffectiveFluid}
  \frac{8\pi G}{3}\bkr_\mathrm{eff}=\sum_a\Fd^{(a)}-\frac{1}{6}\left(\Rd+\Qd^T\right), \quad
  \frac{8\pi G}{3}\bkp_\mathrm{eff}=\frac{1}{3}\frac{6\sum_a\Fd^{(a)}+\Rd-4\Pd-4\BPd-3\Qd^T}{6},
\ee
with the effective equation of state
\be
  w_\mathrm{eff}=\frac{\bkp_\mathrm{eff}}{\bkr_\mathrm{eff}}=-\frac{1}{3}\frac{\Rd-4\Pd-4\BPd-3\Qd^T+6\sum_a\Fd^{(a)}}{\Rd-\Qd^T-6\sum_a\Fd^{(a)}} .
\ee
%We should however, take care with interpretation. Different factors in the lapse can cause complications. This is clearly demonstrated in the next section when the use of conformal time and cosmological perturbation theory introduces powers of the background scale factor in the densities and correction terms, but the issue arises more generally. Likewise, in our approach some subsidiary steps must be taken to separate terms that might appear in the usual Friedmann equations from pure modifications. As an example, the modifications for a perturbed FLRW universe with arbitrary spatial curvature were derived in Paper II, and the curvature correction $\Rd$ contains a term $\mathcal{K}/a^2$, which appears in the background Friedmann equations. As a result of these complications, it seems wise to define the effective fluid in a model-dependent manner.
In some situations we should, however, take care with the interpretation of the results. A true Friedmann model is specified by a matter density, a cosmological constant and a curvature term. Depending on the time coordinate a conformal factor can also appear. Likewise, the corresponding Raychaudhuri equation is specified by a matter density, a matter pressure and a cosmological constant. Any or all of these terms can emerge from averaging the underlying structure: the effective cosmological constant would be a constant term that appears in the effective Friedmann equation, and the effective curvature term would be a term in the Friedmann equation proportional to $1/\dom{a}^2$. The effective fluid might then sensibly be chosen so as to isolate these. Choosing the prescription (\ref{RawEffectiveFluid}) would instead include the effective curvature in the effective fluid.

When dealing with a true inhomogeneous model, this distinction is a matter of convention; the behaviour of the averaged scale factor remains the same. However our approach, as in Papers I and II, will involve specifying a background FLRW model and evaluating the corrections to this model from perturbations. The background Friedmann and Raychaudhuri equations will therefore in general include curvature terms and conformal factors. The averaged Friedmann equations, on the other hand, include these terms within $\Rd$ and $\Pd$. For example, equation (3.9) of Paper II shows $\Rd$ in a curved FLRW universe perturbed to second order. This includes the background term $\mathcal{K}/a^2$. In this context -- comparing the averaged equations to the assumed background equations -- this term is not a correction. In other contexts this would not be an issue, but here it might be better to define the effective fluid in a model-dependent manner.

\section{Backreaction in Perturbation Theory}
\subsection{Basic Results}
\label{PertTheory}
The perturbed Robertson-Walker line element with a scale factor $a=a(\eta)$ can be written as
\be
\label{FLRWLineElement}
  ds^2=a^2(\eta)\left(-(1+2\phi)d\eta^2+2B_id\eta dx^i+\left(\delta_{ij}+2C_{ij}\right)dx^idx^j\right)
\ee
where the perturbation $\phi=\sum_n(1/n!)\phi^{(n)}$ contains contributions from all perturbative orders, with corresponding definitions for $B_i$ and $C_{ij}$. In this work we only consider scalar perturbations, although the results of this section are equally valid for spacetimes containing vector and tensor perturbations. The spatial perturbation can be expanded into a trace and traceless component,
\be
\label{SpatialMetricSeparation}
  C_{ij}=-\psi\delta_{ij}+E_{ij},
\ee
where $E^i_i=0$. In general we will only retain quantities to second-order in the perturbations and neglect all higher-orders.\footnote{This includes terms of the form $\av{\Phi^2}^2$. In principle, these quantities could be large; however, it is clear that, within the confines of perturbation theory, these quantities are approximately of fourth-order in perturbations.} To second order, the perturbations are
\be
  \phi=\phi_{(1)}+\frac{1}{2}\phi_{(2)}, \quad B_i=B_i^{(1)}+\frac{1}{2}B_i^{(2)}, \quad C_{ij}=C_{ij}^{(1)}+\frac{1}{2}C_{ij}^{(2)},
\ee
but except in products we never explicitly expand them. Indices on the perturbations are raised and lowered with the Kronecker delta. Derivatives with respect to the conformal time will be denoted with an overdot and $H=\dot{a}/a$ is the conformal Hubble rate of the underlying model. Directly associating the spatial coordinates in (\ref{FLRWLineElement}) and (\ref{3+1LineElement}) and noting that $\beta^i=h^{ij}\beta_j$ gives
\be
\label{FLRWMetric}
  h_{ij}=a^2\left(\delta_{ij}+2C_{ij}\right), \quad
  h^{ij}=a^{-2}\left(\delta^{ij}-2C^{ij}+4C^{ik}C^j_k\right), \quad
  \beta_i=a^2B_i, \quad
  \beta^i=B^i-2C^{ij}B_j.
\ee
Since $\alpha^2-\beta^i\beta_i=a^2(1+2\phi)$, the lapse is then
\be
\label{alpha}
  \alpha^2=a^2\left(1+2\phi+B^2\right)
\ee
where $B^2=B^iB_i$. Paper II contains some discussion on the use of perturbation theory in backreaction, but some further comments are useful here. In essence, FLRW cosmology involves the use of an implicit ``average'' background upon which perturbations of arbitrary order are defined. When we apply a spatial averaging formalism to a perturbed FLRW cosmology, we are averaging across the perturbed manifold. Here, we are taking an average across a surface defined by perturbations that include second-order terms. This naturally introduces an apparent ambiguity between the Hubble rate $\dot{a}/a$ appearing in the usual background equations
\be
  \label{Background}
  \left(\hub\right)^2=\frac{8\pi G}{3}a^2\bkr+\frac{1}{3}a^2\Lambda, \quad
  \frac{\ddot{a}}{a}=-\frac{4\pi G}{3}a^2\left(\bkr+3\bkp\right)+\frac{1}{3}a^2\Lambda+\left(\hub\right)^2
\ee
and in the averaged equations (\ref{BuchertHub}, \ref{BuchertRay}). However, the Hubble rate $\dom{\dot{a}}/\dom{a}$ is the Hubble rate defined by the time variation of an expanding volume living on the perturbed 3-surface, entirely distinct from the Hubble rate of the underlying model. In this paper, as in Paper I and Paper II, we will define the effects of ``backreaction'' to be the difference between these Hubble rates, and between the acceleration rates $\dom{\ddot{a}}/\dom{a}$ and $\ddot{a}/a$.

From the Ricci scalar on the three-surface, the curvature correction is
\bea
\label{RawCurvature}
 \Rd&=&\av{(1+2\phi)\left(2\pd^i\pd^jC_{ij}-2\pd^i\pd_iC\right)+4C_{ij}\pd^i\pd^jC+4C^{ij}\pd^k\pd_kC_{ij}-8C_{ij}\pd^k\pd^iC^j_k+3\pd^iC^{jk}\pd_iC_{jk}} \nonumber
   \\ &&\quad -\av{2\left(\pd^iC^{jk}\right)\left(\pd_jC_{ik}\right)-\left(\pd^iC\right)\left(2\pd^jC_{ij}-\pd_iC\right)+2\left(\pd_iC^{ij}\right)\left(2\pd^kC_{jk}-\pd_jC\right)} .
\eea
The extrinsic curvature is
\be
  \alpha K^i_j=-\frac{\dot{a}}{a}\delta^i_j-\dot{C}^i_j+\frac{1}{2}\left(\pd^iB_j+\pd_jB^i\right)
    -\left(\pd^iC^k_j+\pd_jC^{ik}-\pd^kC^i_j\right)B_k+2C^{ik}\dot{C}_{jk}-C^{ik}\left(\pd_kB_j+\pd_jB_k\right)
\ee
and so the total kinematical backreaction is
\bea
\label{RawKinematic}
  \Qd^T&=&-4\hub\av{\pd^iB_i-B^i\left(2\pd^jC_{ij}-\pd_iC\right)-2C^{ij}\pd_iB_j}-\frac{2}{3}\av{\dot{C}}^2 \nonumber \\
    &&\quad +\av{\dot{C}^2-\dot{C}^i_j\dot{C}^j_i+\left(\pd^iB_i\right)^2-\frac{1}{2}\left(\pd^iB_j\right)\left(\pd^jB_i\right)
      -\frac{1}{2}\left(\pd^iB_j\right)\left(\pd_iB^j\right)-2\dot{C}\pd^iB_i+2\dot{C}^i_j\pd_iB^j} \, .
\eea
The dynamical backreactions become
\bea
\label{RawDynamical}
  \Pd&=&3\left(\hub\right)^2+\av{\pd^i\pd_i\phi+\hub\left(3\dot{\phi}+\dot{C}\right)-\hub\pd^iB_j} \nonumber \\ &&\quad
    +\av{B_i\pd^j\pd_jB^i+\left(\pd^iB_j\right)\left(\pd_iB^j\right)-\left(\pd^i\phi\right)\left(\pd_i\phi\right)-2C^{ij}\pd_i\pd_j\phi-\left(\pd^i\phi\right)\left(2\pd^jC_{ij}-\pd_iC\right)+\dot{\phi}\left(\dot{C}-\pd^iB_i\right)} \nonumber \\ &&  \quad
    -\hub\av{3B_i\dot{B}^i+6\phi\dot{\phi}+2C^{ij}\dot{C}_{ij}-2C^{ij}\pd_iB_j-B^i\left(2\pd^jC_{ij}-\pd_iC\right)}
\eea
and
\bea
\label{RawBetaP}
  \BPd&=&\av{\pd^i\dot{B}_i+6\hub\pd^iB_i}-\av{2C^{ij}\pd_i\dot{B}_j+2\dot{C}^{ij}\pd_iB_j+12\hub C^{ij}\pd_iB_j+\left(\pd^iB_i\right)^2}
 \nonumber \\ && \quad
  -\av{\dot{B}^i\left(2\pd^jC_{ij}-\pd_iC\right)+B^i\left(2\pd^j\dot{C}_{ij}-\pd_i\dot{C}\right)+6\hub B^i\left(2\pd^bC_{ij}-\pd_iC\right)}
 \\ && \quad
   +\av{B^i\left(\pd_i\dot{C}-\pd_i\pd^j\dot{B}_j\right)+2\dot{C}\pd^iB_i-3\hub B^i\pd_i\phi} . \nonumber
\eea
The dynamical backreaction then includes a background term $3(\dot{a}/a)^2$ which appears in the background Einstein equations (\ref{Background}). To ensure that the ``backreaction fluid'' contains only correction terms, we include this in the averaged Raychaudhuri equation separately from $\Pd$. Since there is no gauge choice in which $B^a=C^{ab}=0$, from (\ref{RawCurvature}, \ref{RawKinematic}, \ref{RawDynamical}, \ref{RawBetaP}) it is already apparent that there exists no gauge in which all the correction terms vanish.

Finally, we have the fluid corrections. For a perfect fluid with $T^{(a)}_{\mu\nu}=(\rho_{(a)}+p_{(a)})u^{(a)}_\mu u^{(a)}_\nu+p_{(a)} g_{\mu\nu}$, the correction term $\Fd^{(a)}$ is
\be
  \Fd^{(a)}=\frac{8\pi G}{3}\av{\left(\rho_{(a)}+p_{(a)}\right)\left(\left(\alpha n^\mu u^{(a)}_\mu\right)^2-\alpha^2\right)}
     =\frac{8\pi G}{3}\av{\left(\rho_{(a)}+p_{(a)}\right)\left(\left(u^{(a)}_0-\beta^iu^{(a)}_i\right)^2-\alpha^2\right)} .
\ee
The rest-frame energy density and pressure are expanded to second-order as
\be
\label{Linearise}
  \rho_{(a)}=\bkr_{(a)}+\delta\rho_{(a)}=\bkr_{(a)} (1+\delta_{(a)}), \quad p_{(a)}=\bkp_{(a)}+\delta p_{(a)} .
\ee
Here $\bkp_{(a)}=w_{(a)}\bkr_{(a)}$, and if we define the speed of sound to be $c_{s,(a)}^2=\partial p_{(a)}/\partial\rho_{(a)}$ then the second-order expansion of the pressure perturbation is
\be
  \delta p_{(a)}=\frac{\partial p_{(a)}}{\partial\rho_{(a)}}\delta\rho_{(a)}+\frac{1}{2}\frac{\partial^2p_{(a)}}{\partial^2\rho_{(a)}}(\delta\rho_{(a)})^2
          =\bkr_{(a)}\left(c_{s,(a)}^2\delta_{(a)}-\frac{(c_{s,(a)}^2)^\cdot}{2(1+w_{(a)})H}\delta_{(a)}^2\right)
\ee
where $\delta_{(a)}$ includes contributions from all perturbative orders and we have used the background matter continuity equation. With the 4-velocity
\be
  u^\mu_{(a)}=a^{-1}\left(1-\phi+\frac{1}{2}\left(v_{(a)}^2+3\phi^2-2B_iv_{(a)}^i\right),v_{(a)}^i\right)
\ee
where $v_{(a)}^2=\delta_{ij}v_{(a)}^iv_{(a)}^j$, $\Fd^{(a)}$ becomes
\be
  \Fd^{(a)}=\frac{8\pi G}{3}a^2\left(\bkr_{(a)}+\bkp_{(a)}\right)\av{\left(v^{(a)}_i+B^{(a)}_i\right)\left(v_{(a)}^i+B_{(a)}^i\right)} .
\ee
Expanding $\av{\alpha^2\rho_{(a)}}$ and $\av{\alpha^2p_{(a)}}$ with (\ref{alpha}) and (\ref{Linearise}), we can finally express the Buchert equations in a perturbed FLRW universe in an unfixed gauge as
\bea
  \left(\Hd\right)^2&=&\frac{8\pi G}{3}a^2\sum_a\bkr_{(a)}+\frac{1}{3}a^2\Lambda-\frac{1}{6}\left(\Rd+\Qd^T-6\sum_a\Td^{(a)}-6\Ld\right), \\
  \left(\frac{\dom{\ddot{a}}}{\dom{a}}\right)&=&-\frac{4\pi G}{3}a^2\sum_a\left(\bkr_{(a)}+3\bkp_{(a)}\right)+\frac{1}{3}a^2\Lambda+\left(\hub\right)^2
\nonumber \\ && \qquad
    +\frac{1}{3}\left(\Pd+\BPd+\Qd^T+3\Ld-3\sum_a\left(\frac{1}{2}\Td^{(a)}+\Sd^{(a)}\right)\right)
\eea
where now
\bea
  \Td^{(a)}&=&\frac{8\pi G}{3}a^2\bkr_{(a)}\av{\delta_{(a)}+2\phi+2\phi\delta_{(a)}+B^2+(1+w_{(a)})\left(v^{(a)}_i+B_i\right)\left(v_{(a)}^i+B^i\right)}, \\
  \Sd^{(a)}&=&\frac{4\pi G}{3}a^2\bkr_{(a)}\av{3c_s^2\delta_{(a)}+6w\phi+6c_{s,(a)}^2\phi\delta_{(a)}+3w_{(a)} B^2
\right. \nonumber \\ && \qquad \left.
    +(1+w_{(a)})\left(v^{(a)}_i+B^{(a)}_i\right)\left(v_{(a)}^i+B_{(a)}^i\right)-\frac{(c_{s,(a)}^2)^\cdot}{2(1+w_{(a)})H}\delta_{(a)}^2}
\eea
correct the energy density and pressure respectively and $\Ld$ is a correction dependent on the cosmological constant,
\be
  \Ld=\frac{1}{3}a^2\Lambda\av{2\phi+B^2} .
\ee
The effective fluid giving the differences between the base and re-averaged FLRW model then has the effective energy density and pressure
\be
  \frac{8\pi G}{3}a^2\bkr_\mathrm{eff}=\frac{6\sum_a\Td^{(a)}+6\Ld-\Rd-\Qd^T}{6},
\ee
\be
  \frac{8\pi G}{3}a^2\bkp_\mathrm{eff}=\frac{1}{3}\frac{\Rd-4\Pd-4\BPd-3\Qd^T+12\sum_a\Sd^{(a)}-18\Ld}{6}
\ee
and an effective equation of state
\be
  \weff=-\frac{1}{3}\frac{\Rd-4\Pd-4\BPd-3\Qd^T+12\sum_a\Sd^{(a)}-18\Ld}{\Rd+\Qd^T-6\sum_a\Td^{(a)}-6\Ld} .
\ee
These can be compared with equations (33-35) of Paper II, generalised to a non-vanishing shift vector and conformal time.

\subsection{Gauge Choices}
\label{Gauges}
The equations presented thus far are in an entirely unfixed gauge and can be adapted with relative ease to any perturbed FLRW universe. From the form of the correction terms in general, it is immediately apparent that we cannot find a gauge in which each contribution vanishes. The remaining question is whether we can find a gauge in which the corrections take on the simplest possible form. The most obvious choices are those gauges employed in standard cosmological perturbation theory for their simplicity: the uniform curvature, or flat, gauge; the uniform density gauges; the conformal Newtonian gauge; and the conformal synchronous gauge. Uniform curvature gauge would remove the curvature correction, a uniform density gauge should simplify the fluid corrections, the conformal Newtonian gauge should simplify the anisotropies significantly, and a synchronous gauge simplifies the dynamical backreaction. The gauge transformations and governing equations in FLRW universes perturbed up to second order can be found in \cite{Malik08} and we will not repeat them here.

The uniform curvature gauge is of immediate interest in backreaction. In this gauge the spatial surfaces align with those of the background FLRW model and the inhomogeneities are carried entirely by the choice of threading. As a result, by definition the curvature correction $\Rd$ vanishes in this gauge, although this is balanced against the introduction of the additional dynamical backreaction $\BPd$ and shift terms in the kinematical backreaction. No other trivial choice of gauge entirely removes one of the backreaction terms when the system is written in conformal time; even for conformal synchronous gauge there is a non-zero contribution to $\Pd$ arising from the time-derivative of $\alpha$. A further benefit of the uniform curvature gauge is that for scalar perturbations the metric determinant reduces to $\sqrt{h}=a^3$, implying that the comoving volume of a spatial domain remains constant, which simplifies the averaging procedure significantly.\footnote{In the terminology of \cite{Paranjape09} and other works by these authors, a flat gauge is an example of a ``volume-preserving'' gauge.}

The conformal Newtonian gauge is also easily motivated -- while it doesn't remove any of the individual correction terms, the perturbation to the 3-metric is diagonalised and the shift is vanishing, which in particular considerably simplifies $\Rd$ from the general case. It is also straightforward to find analytical solutions for simple systems in conformal Newtonian gauge, justifying its study (see for example \cite{Durrer04,Green05}).

Another interesting choice might be a uniform density gauge. Here the slicing is defined by choosing $\delta\rho=0$, and the threading by setting for example $E=0$ in (\ref{SpatialMetricSeparation}). In this gauge, then, the velocity, shift and lapse are all non-vanishing. We can immediately see that the curvature correction will take the same form as that in conformal Newtonian gauge but that the kinematical and dynamical backreactions are significantly more complicated than in either of the other two gauges. Furthermore, the aim of a uniform density gauge would be to simplify the fluid corrections $\Td$ and $\Sd$, and this it fails to do. Setting $\delta=0$ leaves the fluid corrections as unwieldy combinations of the lapse, shift and velocity.

The final obvious choice is conformal synchronous gauge. We do not consider this gauge here, since the curvature correction $\Rd$ is complicated (in real space) by the anisotropic components of the metric perturbation, and it has been well studied in the literature before (as for example in \cite{Li07,Li08}). However, conformal synchronous gauge does have the advantage that the scale factor $a_D$ corresponds to a volume containing constant mass and therefore perhaps has a clearer physical meaning than in the other gauges.\footnote{The authors are grateful to Dominik Schwarz for highlighting this point.}

We will hence focus on the two choices that seem most likely to yield simple results: the uniform curvature and the conformal Newtonian gauges. In the current work it is not our aim to solve the resulting equations. Instead, we consider the form of the corrections in two particular, simple gauge choices. We will, however, briefly consider pure dust universes in both cases.

\subsubsection{Uniform Curvature Gauge}
The uniform curvature gauge is of direct interest since the spatial hypersurfaces align with those of the FLRW background. The volume element on the hypersurfaces in flat gauge, and therefore the spatial averages across those surfaces, then coincide with the FLRW background.

The metric in the uniform curvature gauge is found by setting $C_{ab}=0$ in the above. Then
\be
  ds^2=a^2(\eta)\left(-(1+2\phi)d\eta^2+2B_id\eta dx^i+\delta_{ij}dx^idx^j\right)
\ee
and so the spatial metric, shift and lapse are
\be
  h_{ij}=a^2\delta_{ij}, \quad \beta^i=B^i, \quad \alpha^2=a^2(1+2\phi+B^2) .
\ee
This gauge being spatially flat, the curvature correction is trivially
\be
  \Rd=0 .
\ee
The kinematical backreaction (\ref{RawKinematic}) becomes
\be
 \Qd^T=\av{\left(\pd^iB_i\right)^2-\frac{1}{2}\left(\pd^iB_j\right)\left(\pd_iB^j\right)-\frac{1}{2}\left(\pd_jB^i\right)\left(\pd_iB^j\right)-4\hub\pd^iB_i} .
\ee
Subtracting off the Hubble term, the dynamical backreaction (\ref{RawDynamical}, \ref{RawBetaP}) is
\bea
  \Pd&=&\av{\pd^i\pd_i\phi+B^i\pd^j\pd_jB_i+\left(\pd^iB_j\right)\left(\pd_iB^j\right)-\left(\pd_i\phi\right)\left(\pd^i\phi\right)}-\av{\dot{\phi}\pd^iB_i}
   -\hub\av{\pd^iB_i-3\dot{\phi}-3B_i\dot{B}^i+6\phi\dot{\phi}} , \\
  \BPd&=&\av{\pd^i\dot{B}_i-\left(\pd^iB_i\right)^2-B^i\pd_i\pd^jB_j}+3\hub\av{2\pd^iB_i-B^i\pd_i\phi} .
\eea
In flat gauge the fluid velocity takes the form
\be
  u_{(a)}^\mu=\frac{1}{a}\left(1-\phi+\frac{1}{2}\left(v_{(a)}^2+2B_iv_{(a)}^i+3\phi^2\right),v_{(a)}^i\right)
\ee
and so the fluid corrections become
\be
 \Td^{(a)}=\frac{8\pi G}{3}a^2\bkr_{(a)}\av{\delta_{(a)}+2\phi+2\phi\delta_{(a)}+B^2+(1+w_{(a)})\left(v^{(a)}_i+B_i\right)\left(v_{(a)}^i+B^i\right)}
\ee 
and
\bea
 \Sd^{(a)}=\frac{4\pi G}{3}a^2\bkr_{(a)}\av{3c_{s,(a)}^2\delta_{(a)}+6w_{(a)}\phi+6c_{s,(a)}^2\phi\delta_{(a)}+3w_{(a)} B^2}
\nonumber \\ \qquad 
   +\av{\left(1+w_{(a)}\right)\left(v^{(a)}_i+B_i\right)\left(v_{(a)}^i+B^i\right)-\frac{\left(c_{s,(a)}^2\right)^\cdot}{2(1+w_{(a)})H}\delta_{(a)}^2}.
\eea
The correction arising from the cosmological constant is
\be
  \Ld=\frac{1}{3}a^2\Lambda\av{2\phi+B^2} .
\ee

We can now consider a pure dust (or ``Einstein-de Sitter'') universe, in which $w=c_s^2=\Lambda=0$. This case has the dual benefits of being one of the simplest cosmologies, and of being a reasonable approximation to the real universe for a range of redshifts from $z\approx 1000$ to $z\approx 1$. In this universe, the fluid corrections simplify slightly to
\be
 \Td=\frac{8\pi G}{3}a^2\bkr\av{\delta+2\phi+2\phi\delta+B^2+\left(v_i+B_i\right)\left(v^i+B^i\right)}, \quad
 \Sd=\frac{4\pi G}{3}a^2\bkr\av{\left(v_i+B_i\right)\left(v^i+B^i\right)}, \quad
 \Ld=0 .
\ee
Working in Fourier space simplifies these expressions somewhat. Adopting the Fourier convention
\be
\label{Fourier}
  A_\mathbf{k}=\int A(\mathbf{x})e^{i\mathbf{k}\cdot\mathbf{x}}d^3\mathbf{k}, \quad
  A(\mathbf{x})=\int A_\mathbf{k}e^{-i\mathbf{k}\cdot\mathbf{x}}\frac{d^3\mathbf{k}}{(2\pi)^3} ,
\ee
the spatial average of a quantity $A(\mathbf{x})$ can be written as
\be
  \av{A(\mathbf{x})}=a^3\int\dom{W}^*(\mathbf{k})A(\mathbf{k})\frac{d^3\mathbf{k}}{(2\pi)^3}
\ee
where $W(\mathbf{x})$ is a window function defining the domain and $\dom{W}(\mathbf{x})=W(\mathbf{x})/\Vd$. Let $A(\mathbf{x})=f(\mathbf{x})g(\mathbf{x})$. If $f(\mathbf{x})$ and $g(\mathbf{x})$ are both linear perturbations, we can take ensemble averages (denoted with an overbar) of both sides and use the definition of the power spectrum,
\be
  \overline{f(\mathbf{k})g^*(\mathbf{k'})}=\frac{2\pi^2}{k^3}\mathcal{P}(k)\delta(\mathbf{k-k}')
\ee
to find
\be
  \overline{\av{f(\mathbf{x})g(\mathbf{x})}}=\frac{1}{(2\pi)^3}\int\mathcal{P}(k)f(k)g^*(k)\frac{dk}{k}
\ee
where we have used that $a^3\dom{W}(\mathbf{k}=\mathbf{0})=1$. Writing the shift and velocity as
\be
  v_i=i\hat{k}_iv, \quad B_i=i\hat{k}_iB
\ee
therefore allows us to write the kinematical backreaction as
\be
  \Qd^T=-4\hub\overline{\av{\int kB(\mathbf{k})e^{-i\mathbf{k}\cdot{x}}\frac{d^3\mathbf{k}}{(2\pi)^3}}} .
\ee
In uniform curvature gauge, then, the entire kinematical backreaction is carried on averages of pure -- \emph{i.e.}, non-quadratic -- perturbations, and assuming these to vanish as in Papers I and II would unsurprisingly lead to an incomplete answer. This is the most important result of the calculation in this gauge. The other correction terms become
\bea
  \Pd&=&\overline{\av{\int\left(3\hub\dot{\phi}(\mathbf{k})-4\hub kB(\mathbf{k})-k^2\phi(\mathbf{k})\right)
     e^{-i\mathbf{k}\cdot{x}}\frac{d^3\mathbf{k}}{(2\pi)^3}}}
 \nonumber \\ && \quad 
   +\frac{1}{(2\pi)^3}\int\left(\hub\left(3B(k)\dot{B}^*(k)-6\phi(k)\dot{\phi}^*(k)\right)-k\dot{\phi}(k)B^*(k)-k^2\left|\phi(k)\right|^2\right)\mathcal{P}(k)\frac{dk}{k}, \\
  \BPd&=&\overline{\av{\int k\left(\dot{B}(\mathbf{k})+6\hub B(\mathbf{k})\right)e^{-i\mathbf{k}\cdot{x}}\frac{d^3\mathbf{k}}{(2\pi)^3}}}
    +\frac{3}{(2\pi)^3}\hub\int kB(k)\phi^*(k)\mathcal{P}(k)\frac{dk}{k}, \\
  \Td&=&\left(\hub\right)^2\overline{\av{\int\left(\delta(\mathbf{k})+2\phi(\mathbf{k})\right)\frac{d^3\mathbf{k}}{(2\pi)^3}}}
 \nonumber \\ && \quad 
    +\frac{1}{(2\pi)^3}\hub^2\int\left(2\phi(k)\delta^*(k)+2v(k)B^*(k)+2\left|B(k)\right|^2+\left|v(k)\right|^2\right)\mathcal{P}(k)\frac{dk}{k}, \\
  \Sd&=&\frac{1}{2(2\pi)^3}\left(\hub\right)^2\int\left(\left|B(k)\right|^2+\left|v(k)\right|^2+2v(k)B^*(k)\right)\mathcal{P}(k)\frac{dk}{k} .
\eea
While analytic solutions in this gauge can be presented, they lack the clarity of those performed in conformal Newtonian gauge, to which we now turn.

\subsubsection{Conformal Newtonian Gauge}
In Paper I and Paper II, along with for example \cite{Tanaka06}, the Newtonian gauge was employed, as an alternative to the common use in studies of averaging of synchronous gauge. Newtonian gauge has certain advantages: the gravitational potential and spatial curvatures have clear physical interpretations and, moreover, remain small across all the scales studied. The Newtonian gauge also possesses a vanishing shift and a diagonal spatial perturbation, which simplifies the calculations somewhat. Here we use conformal Newtonian gauge. The major advantage in doing so is the simplicity with which solutions may be found -- in Paper II the analytic solutions employed were derived in conformal Newtonian gauge.% It additionally makes it easier to transfer between gauges by incorporating it in a unified framework.

Conformal Newtonian gauge is found by substituting $\phi\rightarrow \Psi$, $B_i\rightarrow 0$ and $C_{ab}\rightarrow-\Phi\delta_{ab}$ in the general expressions, and the line element is
\be
  ds^2=a^2(\eta)\left(-(1+2\Psi)d\eta^2+(1-2\Phi)\delta_{ij}dx^idx^j\right) .
\ee
The kinematical backreaction becomes
\be
  \Qd^T=6\av{\dot{\Phi}^2}.
\ee
The dynamical backreaction is
\be
 \Pd=\av{\pd^i\pd_i\Psi+3\hub\left(\dot{\Psi}-\dot{\Phi}\right)}+\av{2\Phi\pd^i\pd_i\Psi-\left(\pd_i\Psi\right)\left(\pd^i\Psi+\pd^i\Phi\right)}
  -3\av{\dot{\Phi}\dot{\Psi}+2\hub\Psi\dot{\Psi}+2\hub\Phi\dot{\Phi}}
\ee 
and the curvature correction is
\be
  \Rd=4\av{\pd^i\pd_i\Phi}+\av{6\left(\pd^i\Phi\right)\left(\pd_i\Phi\right)+8(\Psi+2\Phi)\pd^i\pd_i\Phi} .
\ee
The fluid corrections are
\be
 \Td^{(a)}=\frac{8\pi G}{3}a^2\bkr_{(a)}\av{\delta_{(a)}+2\Psi+2\Psi\delta_{(a)}+(1+w_{(a)})v_{(a)}^2}
\ee 
and
\be
 \Sd^{(a)}=\frac{4\pi G}{3}a^2\bkr_{(a)}\av{3c_{s,(a)}^2\delta_{(a)}+6w_{(a)}\Psi+6c_{s,(a)}^2\Psi\delta_{(a)}
   +(1+w_{(a)})v_{(a)}^2-\frac{(c_{s,(a)}^2)^\cdot}{2H(1+w_{(a)})}\delta_{(a)}^2} ,
\ee 
and the contribution from the cosmological constant is simply
\be
 \Ld=\frac{2}{3}a^2\Lambda\av{\Psi} .
\ee 
These closely resemble those presented in Paper II, with only $\Pd$ exhibiting minor differences from the change in time coordinate.

In the particular case of a pure Einstein-de Sitter universe, the vanishing anisotropic stress implies that $\Phi=\Psi$ and that $\dot{\Phi}=0$. The fluid corrections become
\be
 \Td=\frac{8\pi G}{3}a^2\bkr\av{\delta+2\Psi+2\Psi\delta+v^2}, \quad
 \Sd=\frac{4\pi G}{3}a^2\bkr\av{v^2}, \quad
 \Ld=0 .
\ee
The kinematical backreaction vanishes, while the dynamical backreaction simplifies slightly to
\be
 \Pd=\av{\pd^i\pd_i\Psi}+\av{2\Phi\pd^i\pd_i\Psi-\left(\pd_i\Psi\right)\left(\pd^i\Psi+\pd^i\Phi\right)} .
\ee
The solutions to the linear system in Fourier space are given \cite{Durrer04} by
\be
  \Phi_\mathbf{k}=\mathrm{const.}, \quad \delta=-\frac{1}{6}\Phi_\mathbf{k}k^2\eta^2, \quad
  v=\frac{1}{3}\Phi_\mathbf{k}k\eta, \quad a=a_0\left(\frac{\eta}{\eta_0}\right)^2 .
\ee
Noting as a result of (\ref{Fourier}) that
\be
  \av{(\pd^a\Phi)(\pd_a\Phi)}=-\av{\Phi\pd^a\pd_a\Phi},
\ee
one can find using these solutions that the effective energy density, pressure and equation of state of the backreaction from second-order perturbations are
\be
  \frac{8\pi G}{3}a^2\bkr_\mathrm{eff}=\frac{r+19}{9}Q, \quad
  \frac{8\pi G}{3}a^2\bkp_\mathrm{eff}=\frac{1}{27}Q, \quad
  w_\mathrm{eff}=\frac{1}{3(r+19)}
\ee
where
\be
  r=\frac{L}{Q},
\ee
and
\be
  L=\frac{36}{\eta^2}\overline{\int\left(\delta(\mathbf{k})+2\Phi(\mathbf{k})\right)e^{-i\mathbf{k}\cdot\mathbf{x}}\frac{d^3\mathbf{k}}{(2\pi)^3}}
    +6\overline{\int k^2\Phi(\mathbf{k})e^{-i\mathbf{k}\cdot\mathbf{x}}\frac{d^3\mathbf{k}}{(2\pi)^3}}, \quad
  Q=\frac{R}{(2\pi)^3}\int k^2\left|\Phi(k)\right|^2\mathcal{P}(k)\frac{dk}{k}
\ee
are the linear and quadratic contributions respectively. (Recall that $\delta=\delta_{(1)}+(1/2)\delta_{(2)}$, so the ``linear'' contributions include terms that are at second order in perturbation theory.) Here $R=\int W(\mathbf{x})\sqrt{h_0}d^3\mathbf{x}/\int W(\mathbf{x})\sqrt{h}d^3\mathbf{x}$ is the ratio of the spatial volume as evaluated on the background and the perturbed manifold. $r$ then encodes the magnitude of straight averages across the perturbation compared to averages of the perturbations squared. In other words, it encodes the error implicit in the approach of, for example, Papers I and II, where it was assumed that $L\equiv 0$.\footnote{While the second version of this paper was being prepared, the authors of \cite{Clarkson09} studied the backreaction in Newtonian gauge including explicitly the averages of second-order perturbations and so effectively found $r$ explicitly, in the case where ensemble averages of first-order perturbations vanish.} It is not the aim of this study to evaluate the magnitude of $r$; however, some general statements may be made about the backreaction nonetheless.

The effective pressure is positive definite. To accelerate the universe then requires a negative energy density, implying that $r<-19$. The backreaction then acts as a dark energy for $w_\mathrm{eff}>-1/3$, which requires $r>-20$. Therefore, if
\be
  r\in(-20,-19)
\ee
then the backreaction arising in a pure dust universe will act to accelerate the spatial volume -- as a dark energy, albeit one with a negative effective energy density which serves to reduce the Hubble rate. This is a restrictive requirement and argues against a backreaction that acts as a dark energy arising naturally in a pure dust universe.

If instead, as seems likely, the average of linear modes vanishes on large scales due to their Gaussian nature, and the average of second-order modes is of the same order of magnitude as that from quadratic terms, we have
\be
  \mathcal{O}(r)\approx 1 .
\ee
For $r\in[-10,10]$, we find that the effective equation of state lies between
\be
  \frac{1}{87}<w_\mathrm{eff}<\frac{1}{27}
\ee
and in the case when the averages of second-order perturbations can themselves be entirely neglected we recover
\be
  w_\mathrm{eff}=\frac{1}{57}
\ee
as found in Paper II. From this we can conclude that, for any system where $\mathcal{O}(r)\approx 1$, the equation of state of backreaction from perturbations, up to and including second-order modes, is similar to warm dust and approximately
\be
  w_\mathrm{eff}\approx\frac{1}{60}.
\ee
Only for situations where $r\approx -39/2$ does the effective equation of state become such that the backreaction acts as a dark energy. This could be possible, for example, on smaller scales when the averages over linear modes cannot be neglected.

\section{Discussion}
In this paper we have formulated the averaged cosmological equations in the most general case for which the 3+1 separation holds and considered the nonlinear regime of perturbation theory in an unfixed and in two simple gauges. The formalism we have derived is entirely general for any spacetime in any coordinate system, and with any combination of fluids. This formalism differs from that recently presented in \cite{Larena09} as we have defined the Hubble rate from the change of a volume defined on an arbitrary surface, while the author of \cite{Larena09} defined his Hubble rate from the expansion scalar. Before specialising to a perturbative system, we also highlighted some issues with the interpretation of the ``backreaction'', depending on the choice of underlying system and of the lapse. These issues are illustrated with reference to perturbative models: in such systems the backreaction terms will in general include terms that otherwise appear in the background Friedmann equations; and for systems in a conformal time the energy densities should all be scaled by the conformal factor. It is therefore sensible to suggest that general statements concerning the ``effective fluid'' should be made with care, and that this effective fluid is best defined on a model-specific basis.

One of the motivations for the development of this formalism, other than future study into fully general systems, was to consider the gauge issue in backreaction and, in particular, whether a choice can be found that renders the correction terms in their simplest form. To this end, we applied the general formalism to a Robertson-Walker universe perturbed to second-order in metric and fluid perturbations, working in an unspecified gauge. The resulting averaged equations can then be put into a required gauge by removing the redundant degrees of freedom. We then argued that the most obvious gauge choices are those that either remove or maximally simplify one or more of the backreaction quantities, with the simplest choices being the uniform curvature gauge, the conformal synchronous gauge, and the conformal Newtonian gauge. The system in uniform curvature gauge remains somewhat unwieldy despite the vanishing spatial curvature, with the kinematical and dynamical backreactions complicated instead with contributions from the shift vector. However, flat gauge does significantly simplify the use of Fourier modes and ensemble averaging. The system appears to reduce to a simpler form in the conformal Newtonian gauge, in which the anisotropic terms and the shift vanish and in which analytic solutions for simple cases are easily found. Working in conformal Newtonian gauge, we generalised the analytical estimates of Paper II to the case where contributions of the form $\av{\Phi}$ cannot be neglected, finding that, on large scales where the average of linear modes vanishes and that of second-order modes is of the same order as quadratic terms, the backreaction has an equation of state similar to warm dust. On smaller scales where the average of linear modes no longer vanishes it remains possible to argue for an equation of state more drastically different from dust, but the linear contributions would need to be in a narrow window more than an order of magnitude greater than those from quadratic terms.

\begin{acknowledgments}
The authors wish to thank Christof Wetterich, Thomas Dent, Sami Nurmi, David Seery and Kishore Ananda for helpful discussions and comments. IAB is supported by the Heidelberg Graduate School for Fundamental Physics and wishes to thank the Department of Mathematics and Applied Mathematics in Cape Town for kind hospitality, and Chris Clarkson, George Ellis, Julien Larena and Roy Maartens for stimulating discussions during the stay. They also wish to thank an anonymous referee for comments that helped increase the clarity of the paper. JB and IAB also wish to thank the Astronomy Unit at Queen Mary University of London for kind hospitality during the opening and closing stages of the project.
\end{acknowledgments}

\bibliography{Backreaction}

\end{document}